\documentclass[12pt,a4paper,emulatestandardclasses,openany]{scrreprt}
\usepackage[total={6in, 10in}]{geometry}
\usepackage{latexsym,amsfonts,amssymb,mathrsfs,amssymb,graphicx,authblk,booktabs,multirow}
\usepackage[frozencache=true,cachedir=.]{minted}
\usepackage[table]{xcolor}
\usepackage{hyperref}
\newcounter{Example}
\definecolor{mplc0}{HTML}{1f77b4}
\definecolor{mplc1}{HTML}{ff7f0e}
\definecolor{mplc2}{HTML}{2e7d32}
\definecolor{mplc3}{HTML}{d32f2f}
\hypersetup{
	colorlinks=true,
	linkcolor=mplc0,
	filecolor=mplc1,
	urlcolor=mplc3,
	citecolor=mplc2
}

\title{Job submission on grid}
\author{Rudra Banerjee}
\affil{Advanced Computing Lab\\Department of Physics\\University of Pune}
\begin{document}
\maketitle
\tableofcontents
\chapter*{Preface}
\addcontentsline{toc}{chapter}{Preface}

This is an user's introduction to grid using Globus Toolkit from an user's point of
view. With a brief introduction to what grid is, I have shifted quickly to the game itself. In this
part, i have done an step by step introduction, starting from the access to the grid to submitting
the job.

In the appendix, a special note is there on using GARUDA grid.

\subsubsection*{About new date on title}
Its not an update. When I first wrote this article, I never thought this will be popular. But, with
lack of time to update the article, I have just reformatted it. Its a cosmetic change, not a real
\textit{makeover}. {\color{red}{User must check the current data and state of GARUDA bu himself.}}

Hope this will be of help for the users.
\chapter{The Garuda Grid }\label{ch1}
\section{What is Grid???}
A computational Grid is an aggregation of heterogeneous and geographically distributed resources
such as computing, storage and special equipments. The aim of the grid is to dynamically share
resources to facilitate optimal utilization. Grid allows us to solve complex problems that could
not be dealt with otherwise. It provides high-speed and seamless access to supercomputers and also
enables creation and management of dynamic virtual collaborations and organizations.

Leading scientist is this field, Dr. I. Foster has given  {\textit{Three Point Checklist}} in this
purpose :
\begin{quote}A Grid is a system that
	\begin{itemize}
		\item coordinates resources that are not subject to centralized control \\{\footnotesize
		      Integrates and coordinates different resources and users that live in different control
		      domain}
		\item using standard, open, general-purpose protocols \\{\footnotesize Protocols such as
		      authentication, authorization and access should be open and standard and open. Otherwise the
		      system is application specific.}and
		\item interfaces to deliver nontrivial qualities of service\\{\footnotesize Should deliver
		      various quality of service in a coordinated way so that the utility of the system is
			      {\textbf{significantly}} greater then the sum of its part}.
	\end{itemize}
	and is accepted as present standard.
\end{quote}
\section{Garuda Grid}
GARUDA is a collaboration of science researchers and experimenters on a nation wide grid of
computational nodes, mass storage and scientific instruments that aims to provide the technological
advances required to enable data and compute intensive science for the 21st century. \par GARUDA
aims at strengthening and advancing scientific and technological excellence in the area of Grid and
Peer-to-Peer technologies. To achieve its objective, GARUDA brings together a critical mass of
well-established researchers, from 48 research laboratories and academic institutions, who have
constructed an ambitious program of activities. (35 Research and Academic Institutions, 11 C-DAC
centers and ERNET, Delhi). Right now four nodes of this Grid are working- Pune, Bangalore, Chennai
and Hyderabad center of C-DAC. To provide high performance computing facilities and services for
the scientific and  research community and for the enterprise, C-DAC had earlier set up the
National PARAM Supercomputing Facility (NPSF) at Pune, housing its PARAM 10000 supercomputer with a
peak computing power of 10 Gflops. In its continuing effort in this direction, C-DAC has
established Center for Tera-scale Supercomputing Facility (CTSF) at its C-DAC Knowledge Park,
Bangalore with the PARAM Padma machine in its Tera-scale configuration. The PARAM Padma at CTSF has
a peak computing power of one Tera-flop.

In this four center, not all of the machines are equivalent. Like Pune is Linux cluster, on the
other hand B'lore has Linux, AIX and Solaris. There main capacities and IP are given in the table:
\begin{table}
	\centering
	\begin{tabular}{llll}\toprule
		\rowcolor{mplc1!20}
		Place                      & Resource             & Number                           & IP                                       \\\midrule
		\multirow{3}{*}{B'lore}    & Linux (Xeon) Cluster & \multirow{3}{*}{5 nodes(10 CPU)} &
		\underline{xn02}.ctsf.cdac.org.in                                                                                               \\
		                           & AIX Cluster          &                                  & \underline{tf34}.ctsf.cdac.org.in        \\
		                           & Solaris Cluster      &                                  & \underline{e1}.ctsf.cdac.org.in          \\
		\rowcolor{mplc0!20}Chennai & Linux Cluster        &                                  &
		\underline{che01}.hardware.cdac.ernet.in                                                                                        \\
		H'bad                      & Linux Cluster        & 5 nodes(10 CPU)                  & \underline{hyd01}.hardware.cdac.ernet.in \\
		\rowcolor{mplc0!20}Pune    & Linux Cluster        & 16+head node                     & \underline{xn00}.npsf.cdac.ernet.in      \\\bottomrule
	\end{tabular}
	\caption{Location and resources in GARUDA GRID}
\end{table}
\begin{figure}[!htb]
	\begin{center}
		\includegraphics[width=.9\textwidth]{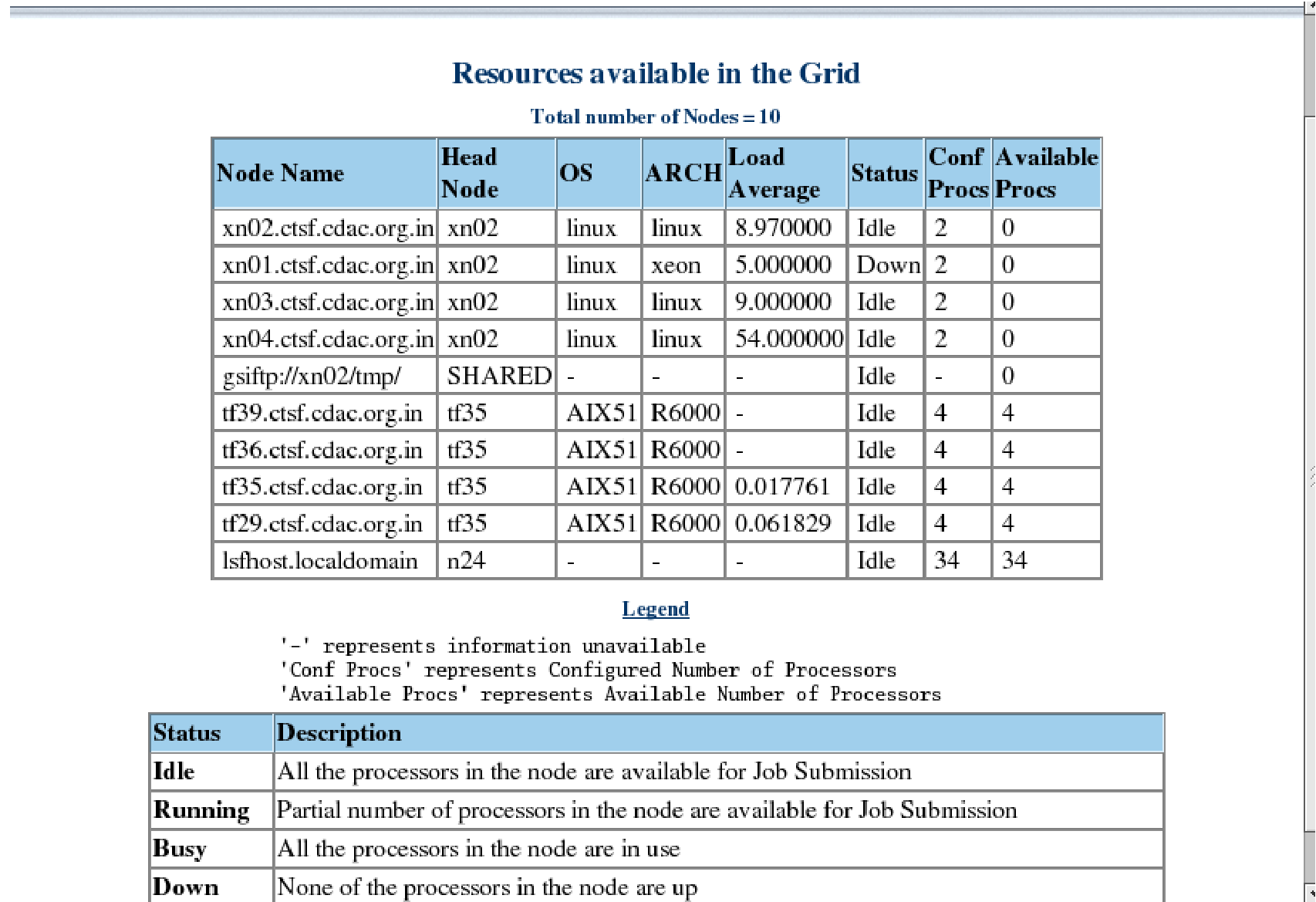}
	\end{center}
	\caption{Grid resources: screen-shot}
\end{figure}
\chapter{Prerequisites}
\section{Getting Permission}First thing first. To submit job in a grid you must have to have a
login and fully working account in the grid-client's node. Next thing is you must have the
\emph{pass phrase} for the grid-proxy to submit jobs in grid. Once you submit
\underline{grid-proxy-init} you will get the access of the grid, for 1 day (24 hr)by default. It is
advisable to get a tentative idea about the running time of your job and if it is more then 1 day,
ask permission for longer period. Some often used option of \underline{grid-proxy-init} are given
below:
\begin{minted}[bgcolor=mplc0!20]{text}
  $grid-proxy-init -h
  Syntax: grid-proxy-init [-help][-pwstdin][-limited][-hours H] ...
  Options
  -help, -usage   Displays usage
  -verify         Verifies certificate to make proxy for
  -pwstdin        Allows passphrase from stdin
  -hours H        Proxy is valid for H hours (default:12)
\end{minted}
\section{Do You Have the Right BINARY File?}
Given that the resource is varied, it is assumed that the user should have a matching binary.
For example, if your binary is compiled on Intel machine, it is most suitable for Xeons; but it is
likely that it may not run on Solaris, as the  Fortarn90 compiler of Solaris is not dependable at
all\footnote{I ll advice to avoid f90 jobs in Solaris as the fortran 90 compiler (CF90)of this
	system is not dependable.} .\par Now few more things you should remember. If you are using a binary
that calls other library file (more often then not, they do) you \emph{static}ally compiled version
of the binary. For Intel fortran, the command is
\begin{minted}[bgcolor=mplc0!20]{text}
  $ ifort -static my file.f90
\end{minted}
\section{Creating binaries in remote machine}
The most important thing in using grid is making the
right binary files for right architecture. Suppose you want to  make a binary for AIX or Solaris.
If you have access to these locally, then there is no problem. But if you dont have, you have to
create it. There's nothing you can do in the RSL to submit a *.f (or any other) and have it be
automatically compiled.

Your main choice (depending on how expensive time on the Solaris machine is) is to just submit a
job to compile the .f file and stage out the executable, then submit that executable.  If you were
confident about compiling it, you could also just submit a shell script that compiled the program
as step one, then ran it as step two.( I would prefer it, as it seems more logical to me and also a
two line script if you are familier with script.)

Otherwise, what you're looking for sounds like cross-compilation.  There was a short thread about
it on the \href{http://gcc.gnu.org/ml/fortran/2006-11/msg00517.html}{gfortran mailing list}.  Basically you can setup
your fortran compiler to emit Solaris executables, but it sounds like more work than just compiling
it directly.

\chapter{A Little on RSL File}
The Globus Resource Specification Language (RSL) provides a neutral way to describe the resource
needs of submitted jobs, a way that can be used by other parts of the Globus job-management system
and that can be translated into each participating computer's local job-management jargon . The
Globus Resource Allocation Manager ({\textit{GRAM}}) relies on RSL strings to perform its management
functions and to coordinate with other Globus software.

RSL provides
\begin{enumerate}
	\item a syntax used to compose complex resource descriptions from basic elements,
	\item a way to record resource decisions as attribute-value pairs for later reuse, and
	\item  a vocabulary of job attributes, each of which serves as a parameter to control job
	      behavior when passed to the (remote) system that actually schedules the job.
\end{enumerate}
The Globus web site offers several detailed descriptions of and technical specifications for RSL.
This subsection (and the next two) summarize the most important RSL features and the aspects of RSL
most relevant for remote users submitting jobs to LC machines

\section{Syntax of RSL Files}

The first recognizer is about job numbers:\\
\begin{tabular}[h]{ll}
	\&  & begins all single-job ({\textit{GRAM}}-managed) resource descriptions. \\
	$+$ & begins all multiple-job (DUROC-managed) resource descriptions.         \\
\end{tabular}

The next thing is different attributes, with \textit{key=val} syntax
\begin{minted}[bgcolor=mplc0!20]{text}
  attribute = "value"
\end{minted}
assigns value to the job resource named by attribute. Each attribute/value pair is called an RSL
relation, and each is separately enclosed in parentheses. On the very first, we generally mention
executable. For my PWscf submission, I have
\noindent{%
	\begin{minted}[bgcolor=mplc0!20]{text}
    (executable=$(GLOBUSRUN_GASS_URL)/home/garuda/rudra/pw/espresso-3.1/bin/pw.x)%
  \end{minted}
}
The next most important attributes are {\texttt{stdin}, \texttt{stdout}} and {\texttt{stderr}}.
They informs \textit{GRAM} about the input, output and the error file. In my above stated PWscf
submission, these three reference are
\begin{minted}[bgcolor=mplc0!20]{text}
  (stdin=$(GLOBUSRUN_GASS_URL)/home/garuda/rudra/al.in)
  (stdout=$(GLOBUSRUN_GASS_URL)/home/garuda/rudra/pwout)
  (stderr=$(GLOBUSRUN_GASS_URL)/home/garuda/rudra/pwerr)
\end{minted}
In above examples, my executable is \texttt{pw.x} which is in the directory
\texttt{/home/garuda/rudra/pw/espresso-3.1/bin/}, my input is \texttt{al.in} which is in
\texttt{/home/garuda/rudra/} and I expect the outputs \texttt{stdout} \& \texttt{stderr} to be in
in the same directory.

The next attribute one need is \emph{arguments}. This is used in various purpose from reading
input values to shifting library files to working site, depending on the argument.

To run PWscf, you need to have your Pseudopotential files in the working node. There is to way to
do that. one is sftp it manually (which is tedious when you need a lot of such file); and the
other is to use \emph{arguments} to do this job. So f my Pseudopotential files are in the
directory \texttt{/home/garuda/rudra/pw/espresso-3.1/Pseudo/} and I need the file \texttt{Al.vbc.UPF}
then the \emph{argument} looks like
\begin{minted}[bgcolor=mplc0!20]{text}
  (arguments=file:///home/garuda/rudra/pw/espresso-3.1/Pseudo/Al.vbc.UPF
  gsiftp://remote\_pwd/Al.vbc.UPF)
\end{minted}
remote pwd is the slot where your job is supposed to run. This is the full {\texttt{rsl}} file, let it
be named as  \texttt{pw.rsl}

\section{Other Commands}
According to your need, you may need many other commands to submit jobs successfully and comfortably.
One of the is already shown in argument. There my file \texttt{\$HOME/pw/espresso-3.1/Pseudo/Al.vbc.UPF}
has to be shifted to working node of the grid. \emph{gsigtp} is used to do this job. If the
situation is such that you have to move your file to a third location, then you can use gsiftp
twice. Suppose you want to move a file from working site to a third site. Then you should use the
command
\begin{minted}[bgcolor=mplc0!20]{text}
(arguments=gsiftp://working node/path/target file
                            gsiftp://remote site/path/destination)}
\end{minted}
\emph{globus-url-copy} or \emph{gridftp} are two other command used for same job.

\chapter{A Little on Grid Variables}
\section{GRAM}
The Globus Toolkit includes a set of service components collectively referred to as the \emph{Grid
	Resource Allocation and Management} ({\textit{GRAM}}). {\textit{GRAM}} simplifies the use of remote systems
by providing a single standard interface for requesting and using remote system resources for the
execution of "jobs". The most common use (and the best supported use) of {\textit{GRAM}} is remote job
submission and control. This is typically used to support distributed computing applications.
\textit{GRAM} is designed to provide a single common protocol for requesting and using remote system
resources, by providing a uniform, flexible interface to local job scheduling systems.
	{\textit{GRAM}}
reduces the number of mechanisms required for using remote resources (such as remote compute
systems). Local systems may use a wide variety of management mechanisms (schedulers, queuing
systems, reservation systems, and control interfaces), but users and application developers need to
learn how to use only one {\textit{GRAM}} to request and use these resources. This connection is
maintained through {\textit{GRAM}}. Both sides need work only with {\textit{GRAM}}, so the number of
interactions and protocols that need to be used are greatly reduced.  {\textit{GRAM}} does not provide
scheduling or resource brokering capabilities.
\section{GASS}\emph{Global Access to Secondary Storage} ({\textit{GASS}})
simplifies the porting and running of applications that use file I/O to the Globus environment. Libraries and utilities are provided to eliminate the need to
\begin{enumerate}\item manually login to sites and ftp files \item install a distributed file system
\end{enumerate}
The APIs are designed to allow reuse of programs that use Unix or standard C I/O with little or no modification.
Currently the ftp and x-gass (GASS server) protocols are supported.
The typical user's view of GASS comprises a simple file access API for opening, closing, and prefetching files. Planned are some RSL extensions for remote cache management.
\begin{enumerate}\item File Access API (to be defined).\item RSL extensions.
\end{enumerate}
\chapter{Submitting Jobs}
So, now you know a lot about grid terminology and ready to submit (at
least I hope) jobs on Grid. Lets have a look on your resources. You have a binary executable and
inputs to run that file, right? So go for shooting the job. There are several options for shooting
jobs on Grid.\section{globusrun} The easiest submission is \emph{globusrun}. The \emph{*.rsl} file
is used here and I think it is best to start with that.
\begin{minted}[bgcolor=mplc0!20]{text}
  $ globusrun -s -r abcd -f pw.rsl
\end{minted}
where \texttt{abcd} is the remote machine where I want to fire the job. Check the table at
Ch.~\ref{ch1} for actual host name.
The globusrun syntax is as follows:
\begin{minted}[bgcolor=mplc0!20]{text}
  Syntax: globusrun [options] [RSL String]
  -help| -usage         Display help
  -i   | -interactive   Run globusrun in interactive mode (multi requests only)
  -f   |-file <file>    Read RSL from the local file <rsl filename>. The RSL
                        can be either a single job request, or a multi request
  -q   |-quiet          Quiet mode (do not print diagnostic messages)
  -o   | -output-enable Use the GASS Server library to redirect standout output
                        and standard error to globusrun. Implies -quiet
  -s   | -server        GLOBUSRUN_GASS_URL can be used to access files local
                        to the submission machine via GASS. Implies
                        -output-enable and -quiet
  -w   | -write-allow   Enable the GASS Server library and allow writing to
                        GASS URLs. Implies -server and -quiet.
\end{minted}

\section{globus-job-run}globus-job-run is an on line interface to job submission, featuring staging of data and executables using a GASS server. Staging is the process in which files are copied from the local machine to the remote machine on which the processing takes place and automatically deleted from it when processing is complete.
This command is used for short, interactive jobs in foreground.\par
The basic syntax is
\begin{minted}[bgcolor=mplc0!20]{text}
  $globus-job-run work-site -s binary arg
\end{minted}
For multiple submission, it looks like:\\
\begin{minted}[bgcolor=mplc0!20]{text}
  $  globus-job-run
  $ -: host1 –np 2 –s myprog.linux arg1
  $ -: host2 –np 3 –s myprog.aix arg2
\end{minted}
where \emph{host} is the remote work node, \emph{-np \textsf{n}} is the number of processors you
want for a job called \emph{myprog.*} .
Here you can see that since I am using same program but compiled in different system(Linux and AIX)
I am supposed to choose my host accordingly.\\Other frequently used options are
\texttt{stdin,stderr {\normalfont \&} stdout}. Job stdin defaults to \texttt{/dev/null}.\\For other
commands, go
for
\begin{minted}[bgcolor=mplc0!20]{text}
  globus-job-run -help
\end{minted}
\section{globus-job-submit}

This is little better then \textit{globus-job-run} in the sense that that it has a batch interface.

The job is submitted using globus-job-submit , its status is then checked using globus-job-status ,
the standard output of the job is the obtained by using globus-job-get-output and then the job is
finally cleaned using the globus-job-clean command.
\appendix
\chapter{The Garuda Portal: How Should You Use It}
There is two way of submitting jobs is Garuda grid.
One is what I talked about...commandline submission; and the second is via Portal. You can access
this only if you are log in to the grid \href{http://192.168.60.40:8080/GridPortal/}{http://192.168.60.40:8080/GridPortal/}.
\par But even you prefer to use command line submission, then also the Portal is very helpful.
Most important of them is monitor the workload on different noads.
\begin{figure}[hbt]\centering
	\includegraphics[width=.75\textwidth]{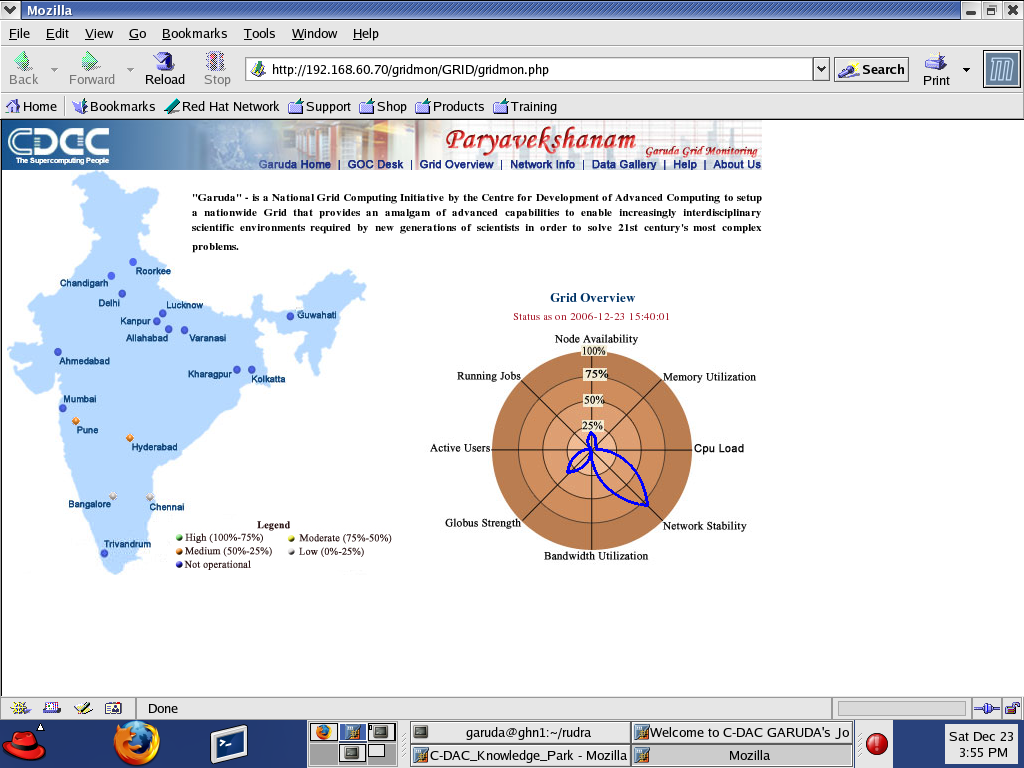}\\
	\includegraphics[width=.75\textwidth]{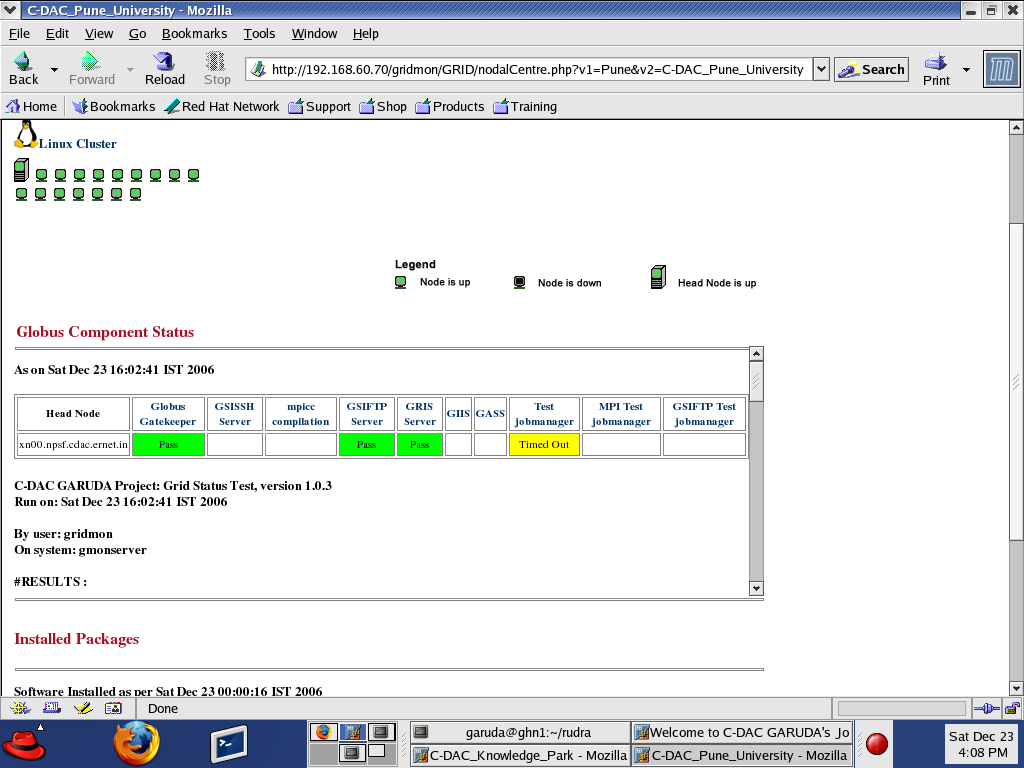}
	\label{garuda}
	\caption{Screen-shot: Garuda Portal}
\end{figure}
By this you can choose where to shoot the jobs. The page shown, gives many first hand information
like CPU load and noad status. From then they give the exact status (Like of there available
resources, status,Installed packages)of different noad.
It is strongly advised to keep a look on the state of a specific node before shooting jobs there.

\end{document}